\documentclass[fleqn,usenatbib]{mnras}

\usepackage[T1]{fontenc}
\usepackage{newtxtext,newtxmath}
\usepackage{graphicx}
\usepackage{amsmath}
\usepackage{xcolor}
\usepackage{comment}

\newcommand{\blue}[1]{#1}

\newcommand{\dd}{\mathrm{d}}

\providecommand{\sovast}{Soviet Ast.}
\providecommand{\aap}{A\&A}
\providecommand{\apj}{ApJ}
\providecommand{\mnras}{MNRAS}
\providecommand{\apss}{Ap\&SS}

\defcitealias{GR2015}{GR}

\title[Massless potential-density pairs]{A radial basis of massless potential-density pairs}

\author[E. V. Polyachenko and I. G. Shukhman]{
Evgeny~V.~Polyachenko$^{1}$\thanks{E-mail: epolyach@inasan.ru}
and Ilia~G.~Shukhman$^{2}$\thanks{E-mail: shukhman@iszf.irk.ru}
\\
$^{1}$Institute of Astronomy, Russian Academy of Sciences, 48 Pyatnitskaya st, Moscow 119017, Russia\\
$^{2}$Institute of Solar-Terrestrial Physics, Russian Academy of Sciences, Siberian Branch, P.O. Box 291, Irkutsk 664033, Russia
}

\date{}
\pubyear{2026}

\begin{document}
\label{firstpage}
\pagerange{\pageref{firstpage}--\pageref{lastpage}}
\maketitle

\begin{abstract}
In the matrix method for linear perturbations of spherical stellar
systems, the perturbed potential and density are expanded over a
set of potential-density pairs. For radial perturbations, mass
conservation makes the coefficient of $1/r$ in the perturbed
potential vanish, so that the potential decays faster than $1/r$.
Starting from the $\ell=0$ Hernquist--Ostriker family, we
construct in closed form a set of potential-density pairs whose
potentials decay as $r^{-p}$ with a prescribed $p\ge2$, the tail
containing all subsequent integer powers. For $p=2$ each pair is
individually massless, and the potential of the $n$-th pair is
expressed through a single Jacobi polynomial. We derive the
combination weights, the leading tail coefficients and the Gram
matrix analytically; the Gram matrix is banded with half-width
$p-1$ (tridiagonal at $p=2$). The $n$-th potential element has
exactly $n-1$ nodes, spread from $r\sim n^{-2}$ to $r\sim n^{2}$,
so that the set resolves both the centre and the far periphery.
As a test, the expansion of the dilation-mode potential of the
isochrone model converges exponentially and carries no parasitic
mass at any truncation, in contrast to the standard set.
\end{abstract}


\section{Introduction}
\label{sec:intro}

The matrix method is a standard tool for studying linear
perturbations of self-gravitating stellar systems
\citep{Kalnajs77, PS81}. The perturbed potential $\Phi(r)$ and
density $\rho(r)$ are expanded over a set of potential-density
pairs $\{\Phi^n,\rho^n\}$ that solve the Poisson
equation pairwise and are biorthonormal, which turns the
linearised Vlasov--Poisson system into a matrix equation for the
expansion coefficients. The quality of the truncated expansion is
controlled by how well a finite number of basis elements can
represent the actual perturbation, and in particular its
behaviour at large radii.

Radial perturbations of stable systems provide the simplest setting for studying damping mechanisms and searching for Landau quasimodes with the matrix method \citep{Wei_94,PSB}.
For radial ($\ell=0$) perturbations the behaviour at infinity is
fixed by mass conservation. The total mass of the system does not
change, so by Gauss's theorem the coefficient of $1/r$ in the
perturbed potential vanishes and $\Phi(r)$ decays faster than
$1/r$. Mass conservation imposes no further constraint: the actual
rate of decay is set by the tail of the perturbed density.

Below
we address one representative case, characteristic of models with
softened point-mass potentials: a tail with the leading power
$r^{-2}$ followed by all subsequent integer powers. A typical example of a radial perturbation is
the potential of the dilation mode \citep{PS23,PS24}, which is an exact stationary solution
that exists in any ergodic model with a single length parameter.
For the isochrone model with potential $\Phi_{\rm iso}(r)=-[1+\sqrt{1+r^2}]^{-1}$ \citep[see][Eq.~4.54]{Hen_60, BT2008} it reads
\begin{equation}
  \Phi_{\rm d}(r)=\frac{1}{d\,(1+d)},\qquad d=\sqrt{1+r^2},
  \label{eq:dil}
\end{equation}
in units $G=M=b=1$, where $M$ is the total mass and $b$ is the  model scale. This function
carries no mass, its leading term at infinity is $1/r^2$, and the
subsequent terms contain \emph{all} integer powers
$1/r^3,\ 1/r^4,\ \dots$ without gaps.
None of the known bases reproduces such a tail. The standard
Clutton--Brock set \citep{CB73} and the recently proposed modified
Clutton--Brock and Even-Power sets \citep{PS26} are all built on the even
variable \[x=\frac{1-r^2}{1+r^2};\] consequently, their large-$r$ expansions
proceed in powers of a single parity: odd powers only for the two
Clutton--Brock bases ($r^{-1}, r^{-3}, r^{-5}, \dots$ for the standard set and
$r^{-3}, r^{-5}, \dots$ for the modified one), and even powers only for the
Even-Power set  ($r^{-2}, r^{-4}, \dots$), built from polynomials in
$x$ without a half-integer power of $1+r^2$ in front. The parity obstruction survives any finite linear
combination of the elements, so a tail containing both parities cannot be represented by any of these sets.

The $\ell=0$ Hernquist--Ostriker
set \citep{HO92} is free of this defect: it is built on the
variable \[\xi=\frac{r-1}{r+1},\] which has no parity in $r$, and its
tails contain all integer powers. It has, however, a defect of
its own: every element decays as $1/r$, that is, carries mass.
Expanding a mass-free function over such elements forces an exact
mutual cancellation of the masses of all elements. At any finite
truncation the cancellation is incomplete, the partial sum
carries a parasitic mass, and the residual weighted with $r^2$
grows linearly with radius, so the tail is never reproduced
uniformly (quantitative results are given in Sect.~\ref{sec:dil}). In the
catalogue of known biorthonormal families \citep{Lil2018} there
is no set whose elements individually decay faster than $1/r$;
this gap was pointed out in \citet{PS24} in connection with the
dilation-mode test of the matrix method.

In this paper we fill the gap. Combining $p$ adjacent elements of
the Hernquist--Ostriker family, we cancel the first $p-1$ inverse
powers of the tail and obtain a set of potential-density pairs whose
potentials decay as $r^{-p}$ with a prescribed $p\ge2$, all subsequent powers
being present. All the required quantities are obtained in closed form: the
combination weights, the leading tail coefficients, and the Gram
matrix, which is banded with half-width $p-1$ and requires no
numerical integration. The case of practical importance is
$p=2$, the tail of the radial perturbations described above. For
$p=2$ the construction is especially simple (one combination coefficient
$K_n$ in closed form, tridiagonal Gram matrix, condition
number growing only as $N^2$ with the number of elements $N$),
and each pair is individually
massless, so that no truncation carries parasitic mass.

The paper is organised as follows. Section~\ref{sec:ho} fixes the
conventions and recalls the Hernquist--Ostriker family in the form
we need. Section~\ref{sec:p2} presents the modified set for
$p=2$. Section~\ref{sec:nodes} describes the number and the
location of the nodes of the new elements, including asymptotic
formulas for the first and the last node. Section~\ref{sec:dil}
tests the set on the expansion of the dilation-mode potential
(\ref{eq:dil}). Section~\ref{sec:concl} summarises the
advantages of the new basis. The construction for arbitrary
$p$ is given in Appendix~\ref{app:p}.

\section{Conventions and the Hernquist--Ostriker family}
\label{sec:ho}

Potential-density pairs are related by the radial Poisson
equation written without the factor $4\pi$,
\begin{equation}
  \left(\frac{\dd^2}{\dd r^2}+\frac{2}{r}\frac{\dd}{\dd r}\right)
  \Phi^n(r)=\rho^n(r),
  \label{eq:pois}
\end{equation}
and are biorthonormal,
\begin{equation}
  \int_0^\infty \Phi^n(r)\,\rho^m(r)\,r^2\,\dd r
  =-\delta_{nm},
  \label{eq:ortho}
\end{equation}
with elements numbered from unity, $n=1,2,3,\dots$ 
The perturbed potential $\Phi$ and density $\rho$, related by the Poisson equation with the factor, $\nabla^2\Phi=4\pi\rho$ at $G=1$
are expanded as
\begin{equation}
\begin{aligned}
  &\Phi(r)=\sum_{n\ge1}C^n\Phi^n(r),\qquad
  \rho(r)=\frac{1}{4\pi}\sum_{n\ge1}C^n\rho^n(r),\\
  &C^n=-\int_0^\infty \Phi(r)\,\rho^n(r)\,r^2\,\dd r =-4\pi\int_0^\infty\Phi^n(r)\,\rho(r)\,r^2\dd r.
\end{aligned}
  \label{eq:coef}
\end{equation}
These are the conventions of \citet{PS81,PS2015,PS24}.

The basis scale is fixed at unity. A free scale
parameter would be redundant: rescaling the basis is equivalent to
rescaling the model under study, since the two scales enter the
problem only through their ratio; if an expansion converges
poorly, the model, rather than the basis, should be rescaled.

The starting point is the $\ell=0$ family of \citet{HO92},
brought to the biorthonormal form
(\ref{eq:ortho}):
\begin{equation}
\begin{aligned}
  \Phi^n(r) &= -\frac{2\sqrt{2n+1}}{n(n+1)}\,
              \frac{C^{(3/2)}_{n-1}(\xi)}{1+r},\\
  \rho^n(r) &= \frac{2\sqrt{2n+1}\;C^{(3/2)}_{n-1}(\xi)}{r\,(1+r)^3},
  \qquad n=1,2,3,\dots,
\end{aligned}
  \label{eq:pair}
\end{equation}
where $C^{(3/2)}_k$ are Gegenbauer polynomials \citep[see, e.g.,][hereafter \blue{GR}]{GR2015}  of the variable
\begin{equation}
  \xi=\frac{r-1}{r+1}.
  \label{eq:vars}
\end{equation}
Biorthogonality (\ref{eq:ortho}) of the pair (\ref{eq:pair}) follows from the orthogonality of the Gegenbauer polynomials \citepalias[][\blue{Eq.~7.313}]{GR2015}.

The construction rests on the following property: $\Phi^n$ is a polynomial in
$y\equiv 1/(1+r)=\frac{1}{2}\,(1-\xi)$ of degree $n$ without a
free term,
\begin{equation}
  \Phi^n=-\sum_{k=0}^{n-1}\beta_{nk}\,y^{k+1},
  \label{eq:polyv}
\end{equation}
with coefficients in closed form,
\begin{equation}
  \beta_{nk}=(-1)^k\,\sqrt{2n+1}\;
     \frac{(n-1)!\,(n+1+k)!}{(n-1-k)!\,(n+1)!\,(k+1)!\,k!}
  \label{eq:beta}
\end{equation}
for $0\le k\le n-1$, and $\beta_{nk}=0$ for $k>n-1$; in particular
\begin{equation}
  \beta_{n0}=\sqrt{2n+1},\qquad
  \beta_{n1}=-\tfrac12\,(n-1)(n+2)\sqrt{2n+1}.
  \label{eq:beta01}
\end{equation}
The first functions are
\begin{equation*}
\begin{aligned}
  \Phi^1&=-\sqrt3\,y,\\
  \Phi^2&=-\sqrt5\,(3y-6y^2),\\
  \Phi^3&=-\sqrt7\,(6y-30y^2+30y^3),\\
  \Phi^4&=-3\,(10y-90y^2+210y^3-140y^4).
\end{aligned}
\end{equation*}
Since
\begin{equation}
  y=\frac{1}{1+r}=\frac{1}{r}-\frac{1}{r^2}+\frac{1}{r^3}-\dots,
\end{equation}
the power $y^k$ produces at infinity the tail $r^{-k}$ followed by
all subsequent integer powers. This is also the origin of the defect of the family: every element carries mass,
\begin{equation}
  \Phi^n(r)\to-\frac{\sqrt{2n+1}}{r},\qquad r\to\infty,
  \label{eq:mass}
\end{equation}
because $\beta_{n0}\ne0$.

\section{The modified set for 
$\lowercase{p}=2$}
\label{sec:p2}

The construction for an arbitrary leading power $p$ is given in
Appendix~\ref{app:p}. Here we present the case needed for radial
perturbations, $p=2$, where a single condition, the vanishing of
the coefficient of $y$ in a combination of two adjacent elements,
$\beta_{n0}-K_n\beta_{n+1,0}=0$, determines the combination
completely:
\begin{equation}
\begin{aligned}
  &\Psi^n=\Phi^n-K_n\,\Phi^{n+1},\qquad
  \sigma^n=\rho^n-K_n\,\rho^{n+1},\\
  &K_n=\sqrt{\frac{2n+1}{2n+3}}
\end{aligned}
  \label{eq:p2}
\end{equation}
The coefficient $K_n$ has a simple interpretation: by
(\ref{eq:mass}) it is the ratio of the masses of two adjacent
elements, precisely the factor that removes the $1/r$ term of the
$n$-th function. No additional normalisation is introduced: the
set follows from the standard one uniquely.

\begin{figure*}
\centering
\includegraphics{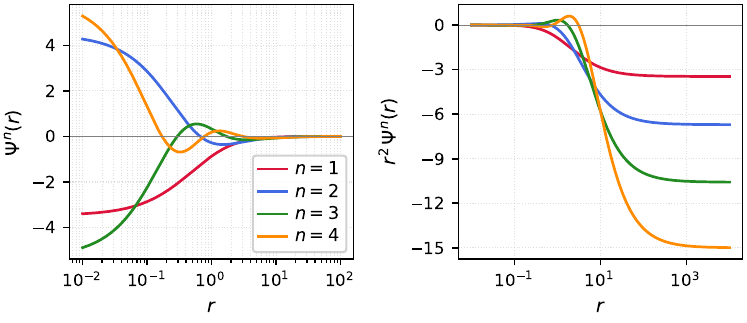}
\caption{The first four elements of the modified set
(\ref{eq:p2}). Left: $\Psi^n(r)$; the element $\Psi^n$ has $n-1$
nodes (Sect.~\ref{sec:nodes}). Right: $r^2\Psi^n(r)$; the curves
approach the constants $c_n=-(n+1)\sqrt{2n+1}$ of
Eq.~(\ref{eq:cn2}), so each element decays exactly as $r^{-2}$
and carries no mass.}
\label{fig:funcs}
\end{figure*}

By construction the coefficient of $y$ in $\Psi^n$ vanishes, so
$\Psi^n$ is $y^2$ times a polynomial of degree $n-1$, and this
polynomial is a Jacobi polynomial. Converting to the Jacobi
polynomials, $C^{(3/2)}_{m}=\frac12\,(m+2)\,P^{(1,1)}_{m}$
\citepalias[\blue{Eq.~8.962.4}]{GR2015}, the combination (\ref{eq:p2}) is
proportional to
\[y\,\Bigl[P^{(1,1)}_{n-1}(\xi)-\dfrac{n}{n+1}\,P^{(1,1)}_{n}(\xi)\Bigr],\]
and the recurrence
relation
\begin{equation}
  (1-\xi)\,P^{(2,1)}_{n-1}(\xi)
  =P^{(1,1)}_{n-1}(\xi)-\frac{n}{n+1}\,P^{(1,1)}_{n}(\xi)
  \label{eq:contig}
\end{equation}
\citepalias[\blue{Eq.~8.961.5}]{GR2015}, together with $1-\xi=2y$, reduces it to a single Jacobi polynomial:
\begin{equation}
  \Psi^n(r)=-\frac{2\sqrt{2n+1}}{n}\,
  \frac{P^{(2,1)}_{n-1}(\xi)}{(1+r)^{2}}.
  \label{eq:jac}
\end{equation}
Each element is therefore \emph{individually}
massless, decays as
\begin{equation}
  \Psi^n(r)\to\frac{c_n}{r^{2}},\qquad
  c_n=-(n+1)\sqrt{2n+1},
  \label{eq:cn2}
\end{equation}
and its tail contains all subsequent integer powers; the value
$P^{(2,1)}_{n-1}(1)=\frac{1}{2}\,n(n+1)$ \citepalias[][\blue{Eq.~8.960}]{GR2015} in (\ref{eq:jac}) reproduces
$c_n$.
The first four
elements are shown in Fig.~\ref{fig:funcs}; the right panel
displays $r^2\Psi^n$ approaching the constants $c_n$.

The Gram matrix $\Delta_{nm}$ follows from the biorthonormality of the parent
family without any integration:
\begin{equation}
\begin{aligned}
  \Delta_{nm}&=-\int_0^\infty\Psi^n\sigma^m r^2\,\dd r
  =\int_0^\infty\frac{\dd\Psi^n}{\dd r}\,
  \frac{\dd\Psi^m}{\dd r}\,r^2\,\dd r,\\
  \Delta_{nn}&=1+K_n^2=\frac{4n+4}{2n+3},\qquad
  \Delta_{n,n+1}=\Delta_{n+1,n}=-K_n,
\end{aligned}
  \label{eq:gram2}
\end{equation}
all other elements being zero: the matrix is tridiagonal.
The modified
set is thus \emph{not} biorthonormal: the price of the prescribed tail is the replacement of the unit Gram matrix by the tridiagonal matrix (\ref{eq:gram2}), so that for a truncation to $N$
elements the expansion coefficients $C^n$, $n=1,\dots,N$, are
found from the system $\sum_{n=1}^{N}\Delta_{mn}C^n=u^m$,
$u^m=-\int_0^\infty\Phi\,\sigma^m r^2\dd r$, rather than by direct
projection; with the band structure known in closed form the
extra cost is negligible. The accuracy with which the $C^n$
are recovered is governed by the condition number of $\Delta$: it
bounds the factor by which a relative error in $u^m$ can be
amplified in the solution. Since $\Delta$ is symmetric and
positive definite, its condition number in the spectral norm is
the ratio of the largest eigenvalue to the smallest,
$\mathrm{cond}_2(\Delta)=\lambda_{\rm max}/\lambda_{\rm min}$ \citep[see, e.g.,][]{GVL2013,Press2007,Higham2002}.
For our matrix it is moderate. Since
$K_n\to 1$ for large $n$, the
limiting matrix is
the tridiagonal
Toeplitz matrix $(-1,2,-1)$, whose
eigenvalues $\lambda_k=2\,\bigl\{1-\cos\, \bigl[k\pi/(N+1)\bigr]\bigr\}$,
$k=1,\dots,N$ \citep[see, e.g.,][\blue{Eq.~4}]{Noschese2013},
lie between $\approx\pi^2/(N+1)^2$ and $4$, so
that $\mathrm{cond}_2(\Delta)\sim(2N/\pi)^2$.

\section{Number and location of the nodes}
\label{sec:nodes}

By (\ref{eq:jac}), the nodes of $\Psi^n$ on $0<r<\infty$ are the
zeros of the Jacobi polynomial $P^{(2,1)}_{n-1}$ on $-1<\xi<1$:
there are exactly $n-1$ of them, all simple, by the classical
theorem on the zeros of orthogonal polynomials.

The location of the nodes follows from the angle variable.
Setting $\xi=\cos\varphi$ in (\ref{eq:vars}) gives
\begin{equation}
  r=\cot^2(\varphi/2),
  \label{eq:angle}
\end{equation}
with the periphery at $\varphi\to0$ and the centre at
$\varphi\to\pi$. The zeros of $P^{(2,1)}_{n-1}(\cos\varphi)$ are 
asymptotically almost equidistant in $\varphi$ (step is $\pi/n+{\cal O}(1/n^2)$) \citep[\blue{theorem 8.21.8}]{Szego75}; the map
(\ref{eq:angle}) is quadratic near both ends, which spreads the nodes over four decades in radius already at moderate $n$.

\begin{figure*}
\centering
\includegraphics{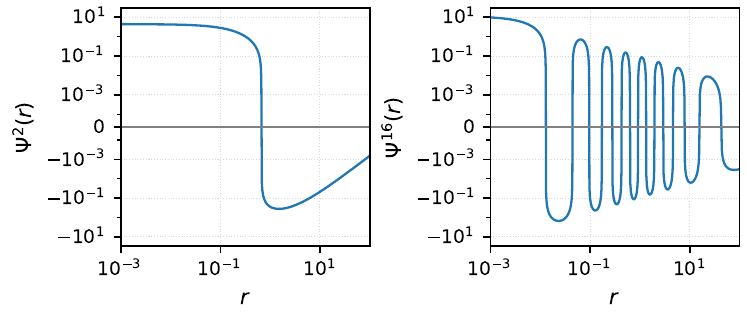}
\caption{Elements $\Psi^2$ (left) and $\Psi^{16}$ (right). The
vertical scale is the standard symmetric logarithmic
(``symlog'') one, linear between $\pm10^{-4}$ and logarithmic
outside, so that the shape of the function and every zero
crossing are visible simultaneously: $\Psi^n$ has exactly $n-1$ nodes, and
for $n=16$ they extend from $r_{\min}=0.013$ to $r_{\max}=43$.}
\label{fig:nodes}
\end{figure*}

The extreme nodes follow from the behaviour of the Jacobi
polynomials near the ends of the interval, where they reduce to
Bessel functions:
\begin{equation}
\begin{aligned}
  P^{(\alpha,\beta)}_{m}(\cos\varphi)={}&
  \frac{(m+\alpha)!}{m!}
  \left(\sin\frac{\varphi}{2}\right)^{\!-\alpha}
  \left(\cos\frac{\varphi}{2}\right)^{\!-\beta}\\
  &\times\left(\frac{\varphi}{\sin\varphi}\right)^{\!1/2}
  \frac{J_\alpha(\varrho\varphi)}{\varrho^{\alpha}}
  +O(m^{-3/2}),
\end{aligned}
  \label{eq:hilb}
  \end{equation}
  where
  \begin{equation}
  \varrho=m+\frac{\alpha+\beta+1}{2},
  \end{equation}
uniformly in $0<\varphi\le\pi-\delta$
\citep[\blue{theorem 8.21.12}]{Szego75}.
The
zeros of
$P^{(\alpha,\beta)}_{m}(\cos\varphi)$ closest to $\varphi=0$
are therefore located at the zeros $j_{\alpha,k}$ of $J_\alpha$,
\begin{equation}
  \varphi_k\simeq\frac{j_{\alpha,k}}{\varrho}\,,
  \label{eq:zeros}
\end{equation}
with relative error $O(m^{-2})$ \citep[\blue{sect.~8.9}]{Szego75}; the
end $\varphi=\pi$ is found using  the reflection formula
$P^{(\alpha,\beta)}_{m}(-\xi)=(-1)^{m}P^{(\beta,\alpha)}_{m}(\xi)$ \citepalias[][\blue{Eq.~8.961.1}]{GR2015},
which swaps $\alpha$ and $\beta$. For our element (\ref{eq:jac}),
$\alpha=2$, $\beta=1$, $m=n-1$, and the order parameter is the
same at both ends, $\varrho=(n-1)+2=n+1$. Hence the extreme
nodes:
\begin{itemize}
\item near the periphery ($\varphi\to0$, index $\alpha=2$) the
element oscillates as $J_2\bigl((n+1)\varphi\bigr)$; the index is
$2$ rather than $1$ precisely because of the double zero $y^2$,
that is, because of the mass cancellation; the last node,
$\varphi\simeq j_2/(n+1)$, lies at
\begin{equation}
  r_{\max}\simeq\frac{(2n+2)^{2}}{j_2^{\,2}},
  \quad j_2=5.1356\ \ (\text{first zero of }J_2);
  \label{eq:rmax}
\end{equation}
\item near the centre ($\varphi\to\pi$, reflected index
$\beta=1$) it oscillates as $J_1\bigl((n+1)(\pi-\varphi)\bigr)$;
the first node, $\pi-\varphi\simeq j_1/(n+1)$, lies at
\begin{equation}
  r_{\min}\simeq\frac{j_1^{\,2}}{(2n+2)^{2}},
  \quad j_1=3.8317\ \ (\text{first zero of }J_1).
  \label{eq:rmin}
\end{equation}
\end{itemize}
The product $r_{\min}\,r_{\max}=(j_1/j_2)^2\approx0.56$ is
independent of $n$. 
Both formulas are confirmed numerically.
Both ends are thus quadratic in the index: $r_{\min}\sim n^{-2}$
and $r_{\max}\sim n^{2}$, so the oscillations of the set cover
radii from $\sim n^{-2}$ to $\sim n^{2}$. For comparison, in the
sets built on the even variable $x=(1-r^2)/(1+r^2)$ \citep{PS26} the extreme
nodes scale only as $\sim 1/n$ and $\sim n$.
The new set therefore
resolves the centre and the far periphery considerably better at equal $N$. This is achieved at the expense of the resolution in the main body of the model: the nodes moved towards the ends by the map (\ref{eq:angle}) are removed from the intermediate region. Counted between
$r=0.1$ and $r=10$, an element of index $n$ has about $0.87\,n$ of its nodes in this interval in the even-variable sets against $0.61\,n$ in
the new set. Equal
resolution of the main body therefore requires about $1.4$ times more elements of the new set, that is, 6 to 8 extra elements
at typical truncation orders $N=15$--$20$. Figure~\ref{fig:nodes} illustrates the node distribution for $n=2$ and $n=16$: the oscillations become denser towards the centre, while their amplitude falls off towards the periphery.

\section{Test: expansion of the dilation mode}
\label{sec:dil}

We expand the dilation-mode potential (\ref{eq:dil}) over the new
set: the coefficients are found from the tridiagonal system of
Sect.~\ref{sec:p2} with $\Phi=\Phi_{\rm d}$ in its right-hand
side, and the partial sum is $S_N=\sum_{n\le N}C^n\Psi^n$. Since $r^2\Phi_{\rm d}\to1$, the
natural quality measure is the residual weighted with $r^2$;
Table~\ref{tab:dil} lists its maximum over $0<r\le10^3$ for the
new set ($p=2$), for the standard set ($p=1$), and for the set
with the leading power deliberately mismatched ($p=3$, see
Appendix~\ref{app:p}). The maximum is attained at the right
endpoint of the interval in every case, so that the table lists the far-field value. The last column is the coefficient of $1/r$ in
$S_N-\Phi_{\rm d}$, that is $\lim r(S_N-\Phi_{\rm d})$, for the
standard set. All the entries are computed in high-precision
arithmetic: in double precision the last rows would be affected by round-off errors.

\begin{table}
\centering
\caption{Convergence of the expansion of the dilation-mode
potential (\ref{eq:dil}).}
\label{tab:dil}
\vspace{4pt}
\setlength{\tabcolsep}{4pt}
\begin{tabular}{c|ccc|c}
\hline
 & \multicolumn{3}{c|}{$\max\bigl|r^2(S_N-\Phi_{\rm d})\bigr|$,
   $0<r\le10^3$}
 & coefficient of $1/r$\\
$N$ & $p=1$ & $p=2$ & $p=3$ & $p=1$\\
\hline
10 & $6.9\cdot10^{-1}$ & $9.1\cdot10^{-3}$  & $0.97$ & $+7.4\cdot10^{-4}$\\
12 & $1.4\cdot10^{-1}$ & $2.3\cdot10^{-3}$  & $0.96$ & $-1.5\cdot10^{-4}$\\
14 & $2.7\cdot10^{-2}$ & $5.4\cdot10^{-4}$  & $0.95$ & $+3.0\cdot10^{-5}$\\
16 & $5.0\cdot10^{-3}$ & $1.2\cdot10^{-4}$  & $0.94$ & $-5.8\cdot10^{-6}$\\
18 & $9.2\cdot10^{-4}$ & $2.8\cdot10^{-5}$  & $0.93$ & $+1.1\cdot10^{-6}$\\
20 & $1.7\cdot10^{-4}$ & $6.0\cdot10^{-6}$  & $0.91$ & $-2.1\cdot10^{-7}$\\
22 & $3.0\cdot10^{-5}$ & $1.3\cdot10^{-6}$  & $0.90$ & $+4.0\cdot10^{-8}$\\
24 & $5.3\cdot10^{-6}$ & $2.6\cdot10^{-7}$  & $0.88$ & $-7.4\cdot10^{-9}$\\
26 & $9.3\cdot10^{-7}$ & $5.4\cdot10^{-8}$  & $0.87$ & $+1.4\cdot10^{-9}$\\
28 & $1.6\cdot10^{-7}$ & $1.1\cdot10^{-8}$  & $0.85$ & $-2.5\cdot10^{-10}$\\
30 & $2.7\cdot10^{-8}$ & $2.2\cdot10^{-9}$  & $0.83$ & $+4.6\cdot10^{-11}$\\
32 & $4.6\cdot10^{-9}$ & $4.3\cdot10^{-10}$ & $0.81$ & $-8.5\cdot10^{-12}$\\
34 & $7.7\cdot10^{-10}$& $8.3\cdot10^{-11}$ & $0.79$ & $+1.5\cdot10^{-12}$\\
36 & $1.3\cdot10^{-10}$& $1.6\cdot10^{-11}$ & $0.77$ & $-2.8\cdot10^{-13}$\\
38 & $2.1\cdot10^{-11}$& $3.0\cdot10^{-12}$ & $0.75$ & $+5.1\cdot10^{-14}$\\
40 & $3.3\cdot10^{-12}$& $5.7\cdot10^{-13}$ & $0.73$ & $-9.1\cdot10^{-15}$\\
\hline
\end{tabular}
\end{table}

Three results 
follow from the Table~\ref{tab:dil}.

\emph{Rate.} At $p=2$ the convergence is exponential, and its
asymptotic rate follows from a classical result of approximation
theory \citep[][for a modern exposition see chapter 8 of
\citealt{Trefethen2013}]{Bernstein1912}: the rate of an expansion
in orthogonal polynomials is governed by the largest ellipse with
foci $\pm 1$ that is free of the singularities of the expanded
function. Such ellipses, $|\xi-1|+|\xi+1|=2a=\zeta+\zeta^{-1}$ in
the complex plane of the expansion variable, are named after
Bernstein; here $\zeta > 1$ is the sum of the
semi-major axis $a$ and the semi-minor axis $\sqrt{a^2- 1}=\frac{1}{2}\,(\zeta-\zeta^{-1})$ of the ellipse.

The expansion is polynomial in
$\xi\in[-1,1]$; the branch points $r=\pm i$ of $\Phi_{\rm d}$ map
to $\xi=\pm i$; the largest singularity-free ellipse with foci
$\pm 1$ then has the sum of semiaxes $\zeta=1+\sqrt2$, so the
residual decays asymptotically as $\zeta^{-N}=e^{-0.88N}$: a
factor $1/\zeta\approx0.414$ per added element, or
$\zeta^{-2}\approx0.17$ between successive rows of
Table~\ref{tab:dil}. The row-to-row factors in the table decrease from $0.25$ to $0.19$, approaching the asymptotic value from above.

The standard set is governed by the same
singularities and has the same asymptotic rate, but at equal $N$
its residual is higher by a factor of $76$ at $N=10$ and of $6$ at $N=40$ (the ratio decreases slowly with $N$): the same accuracy requires five additional elements of the standard set at $N=10$ and two at $N=40$.

\emph{Mass.} The difference is qualitative, not merely
quantitative. At $p=1$ the partial sum carries a parasitic mass
that cancels only in the limit: the residual coefficient of $1/r$
is nonzero and decreases with $N$. Because of it,
$r^2(S_N-\Phi_{\rm d})$ grows linearly with $r$, and at large radii
the residual is not small at any finite $N$: at $N=16$ it equals
$5\cdot10^{-3}$ on $r\le10^3$ but reaches $60$ on $r\le10^7$. At
$p=2$ the mass is zero identically, for each element separately,
so that the $1/r$ term is absent identically and the
last column of Table~\ref{tab:dil} is given for the standard set
only. The weighted residual is bounded on the whole half-axis.

\emph{Choice of $p$.} The leading power of the set must coincide
with the leading power of the function being expanded. The column
$p=3$ illustrates the effect of a mismatch: all elements begin with $1/r^3$, the coefficient of $1/r^2$ cannot be reproduced, and the weighted residual remains of order unity at any
$N$.

\medskip

\begin{figure}
\centering
\includegraphics{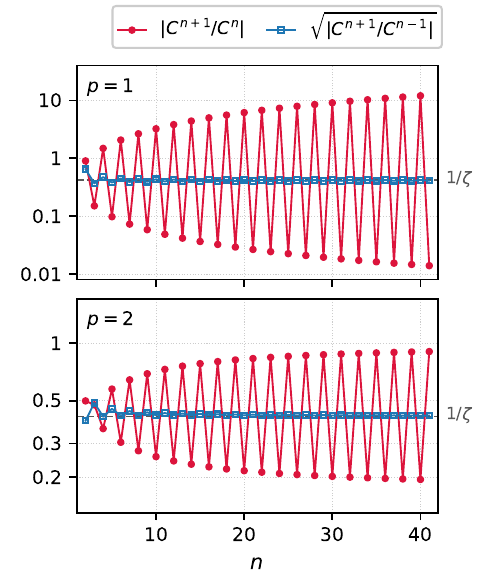}
\caption{Ratios of successive expansion coefficients of the
dilation-mode potential for the standard set ($p=1$, top) and the
new one ($p=2$, bottom). The one-step ratio oscillates with
period two and has no limit; the two-step ratio converges
to $1/\zeta=0.4142$, the rate set by the Bernstein ellipse.}
\label{fig:coef}
\end{figure}

Strictly speaking, the ellipse controls the expansion coefficients, $|C^n|\sim\zeta^{-n}$; the behaviour of the partial sums follows from it. The singularity
$\xi=i$ is the image of the point $z=i\zeta$ of the circle
$|z|=\zeta$ under the Joukowski mapping $\xi=(z+z^{-1})/2$, which
sends this circle to the Bernstein ellipse; the argument of this
point is exactly $\pi/2$ because the singularity is purely
imaginary. The coefficients therefore carry an oscillating
factor, $C^n\propto\zeta^{-n}\cos(n\pi/2+\psi)$: their signs
repeat with period four and their moduli are modulated with
period two. Both effects are seen in Fig.~\ref{fig:coef}. The one-step ratio $|C^{n+1}/C^{n}|$ has no limit: at $p=2$ it oscillates between $0.19$ and $0.90$, and at $p=1$, where
the modulation is far stronger, between $0.015$ and $12$. The modulation cancels over two steps in $n$: the geometric mean of two successive one-step ratios, $\sqrt{|C^{n+1}/C^{n-1}|}$, tends to $0.418$ at $p=2$ against $1/\zeta=0.414$; at $p=1$ it approaches the same limit more slowly, with the two parity branches lying on either side of it. The rate
is a property of the expanded function, and is the same for both
sets.

The partial sums inherit this rate because the residual is
the tail of the series,
$\Phi_{\rm d}-S_N=\sum_{n>N}C^n\Psi^n$.
The elements enter the residual
weighted with $r^2$, and $\max_r|r^2\Psi^n|$ equals the tail
coefficient $|c_n|$ of (\ref{eq:cn2}), which grows only as
$n^{3/2}$; since the algebraic growth is dominated by the geometric decay, the tail is dominated by its first term, $n=N+1$,
that is
\begin{equation}
\begin{aligned}
  \max_{r}\bigl|r^2(S_N-\Phi_{\rm d})\bigr|
  &\simeq\bigl|C^{N+1}c_{N+1}\bigr|\\
  &\propto N^{3/2}\,\zeta^{-N}\,\bigl|\sin(N\pi/2+\psi)\bigr| .
\end{aligned}
  \label{eq:resid}
\end{equation}
The exponential factor sets the rate; the algebraic factor $N^{3/2}$ does not change the exponent but slows the approach to the asymptotic value, which is why the row-to-row factors of Table~\ref{tab:dil}
approach $\zeta^{-2}$ from above; and the modulus of the sine
repeats with period two in $N$, so the modulation is inherited as
well,
but this modulation does not appear in Table~\ref{tab:dil}, since its step in $N$ equals the period of the modulation.

\section{Conclusions}
\label{sec:concl}

Starting from the $\ell=0$ Hernquist--Ostriker family, we have
constructed a set of potential-density pairs whose
potentials decay as $r^{-p}$ with a prescribed $p\ge2$. Its
advantages over the standard sets are the following.

\begin{enumerate}
\item The tail has the physically required form. For radial
perturbations the mass is conserved, the coefficient of $1/r$
vanishes, and the perturbed potential of a model with a softened
point-mass potential decays as $r^{-2}$ with all subsequent
integer powers present. The set with $p=2$ has exactly this
behaviour, element by element; no parity restriction of the
Clutton--Brock type sets and no parasitic mass of the standard
Hernquist--Ostriker set.

\smallskip

\item All quantities are given in closed form. At $p=2$ the combination is
$\Psi^n=\Phi^n-K_n\Phi^{n+1}$ with
$K_n=\sqrt{(2n+1)/(2n+3)}$, which collapses to the single
Jacobi polynomial (\ref{eq:jac}),
$\Psi^n\propto P^{(2,1)}_{n-1}(\xi)\,(1+r)^{-2}$; the tail
coefficients are
$c_n=-(n+1)\sqrt{2n+1}$, and the Gram matrix is tridiagonal with
$\Delta_{nn}=(4n+4)/(2n+3)$, $\Delta_{n,n+1}=-K_n$; no
numerical integration is required at any stage.

\smallskip

\item The conditioning is moderate:
$\mathrm{cond}_2\,\Delta\sim N^2$ at $p=2$.

\smallskip

\item The element $\Psi^n$ has exactly $n-1$ nodes, with the
extreme nodes at $r_{\min}\simeq j_1^2/(2n+2)^2$ and
$r_{\max}\simeq(2n+2)^2/j_2^2$: the oscillations cover radii from
$\sim n^{-2}$ to $\sim n^2$, so the set resolves both the centre
and the far periphery, which the sets built on the even variable, with extreme nodes $\sim n^{-1}$ and $\sim n$ 
do not resolve.

\smallskip

\item In the matrix method the basis enters only through the
values of $\Psi^n(r)$ along the orbits, so the new set can be used in place of the standard one without any change in the algorithm: no formulas beyond (\ref{eq:pair}) and (\ref{eq:p2}) are required.

\smallskip

\item The expansion of the mass-free dilation-mode potential of
the isochrone model converges exponentially, at the asymptotic
rate $e^{-0.88N}$ set by the Bernstein ellipse of its
singularities $r=\pm i$, reaching a given accuracy with two to
five elements fewer than
the standard set, and carries no parasitic mass at any
truncation.

\smallskip

\item The standard set is contained as the special case $p=1$,
and the generalisation to an arbitrary leading power $p$ is
explicit (Appendix~\ref{app:p}).
\end{enumerate}

\section*{Acknowledgements}
This research was partially supported by the Russian Academy of
Sciences Program No.~28 (subprogram II, `Astrophysical Objects as
Cosmic Laboratories') and the Ministry of Science and Higher
Education of the Russian Federation (I.~Shukhman).

\bibliographystyle{mnras}
\bibliography{HO_basis}

\appendix

\section{Arbitrary leading power $p$}
\label{app:p}

\subsection{Construction}

To obtain the leading power $r^{-p}$ one must annihilate the
coefficients of $y,\dots,y^{p-1}$ in (\ref{eq:polyv}). These are
$p-1$ linear conditions, so we combine $p$ adjacent elements:
\begin{equation}
  \Psi^n=\sum_{j=0}^{p-1}w^{(n)}_j\,\Phi^{n+j},
  \qquad
  \sigma^n=\sum_{j=0}^{p-1}w^{(n)}_j\,\rho^{n+j},
  \label{eq:Psi}
\end{equation}
$n=1,2,3,\dots$
The first weight is fixed, $w^{(n)}_0=1$; the rest are determined
uniquely by the system of order $p-1$
\begin{equation}
  \sum_{j=1}^{p-1}\beta_{\,n+j,\,m}\;w^{(n)}_j=-\,\beta_{\,n,\,m},
  \qquad m=0,1,\dots,p-2,
  \label{eq:sysb}
\end{equation}
whose matrix is nondegenerate because the powers
$y,\dots,y^{p-1}$ of $p$ adjacent elements are linearly
independent. At $p=1$
the system is empty, only the weight $w^{(n)}_0=1$ remains, and
$\Psi^n=\Phi^n$: the standard set is a special case of the
construction. At $p=2$ the system
reduces to the single condition of Sect.~\ref{sec:p2}.

Then $\Psi^n=O(y^p)$, that is,
\begin{equation}
  \Psi^n(r)=\frac{c_n}{r^{p}}
            +O\!\left(\frac{1}{r^{p+1}}\right),
  \qquad
  c_n=-\sum_{j=0}^{p-1}w^{(n)}_j\,\beta_{\,n+j,\,p-1},
  \label{eq:cn}
\end{equation}
with all subsequent powers present. 
At $p=2$, Eq.~(\ref{eq:cn}) with $\beta_{n1}$ from
(\ref{eq:beta01}) gives the closed form (\ref{eq:cn2}).

\subsection{Gram matrix}

From the biorthonormality (\ref{eq:ortho}) of the parent family
and the definition (\ref{eq:Psi}),
\begin{equation}
  \Delta_{nm}=-\int_0^\infty\Psi^n\sigma^m r^2\dd r
  =\sum_{j}w^{(n)}_j\,w^{(m)}_{\,j+n-m},
  \label{eq:gram}
\end{equation}
where $0\le j\le p-1$, $0\le j+n-m\le p-1$,
and $\Delta_{nm}=0$ for $|n-m|>p-1$: the matrix is banded with
half-width $p-1$, that is, with $2p-1$ nonzero diagonals. On the
main diagonal,
\begin{equation}
  \Delta_{nn}=1+\sum_{j=1}^{p-1}\bigl(w^{(n)}_j\bigr)^2 .
  \label{eq:gramdiag}
\end{equation}
At $p=1$, when the only weight is $w^{(n)}_0=1$,
Eq.~(\ref{eq:gram}) gives
$\Delta_{nm}=\delta_{nm}$, as required for the standard set.
The condition number 
grows
asymptotically as $N^{2(p-1)}$, that is, as $N^{2}$, $N^{4}$,
$N^{6}$ for $p=2$, $3$, $4$. The exponent has the same origin as
at $p=2$: for $n\to\infty$ the weights tend to the binomial
coefficients, $w^{(n)}_j\to(-1)^j\binom{p-1}{j}$, and the
smallest eigenvalue of the limiting matrix decays as
$N^{-2(p-1)}$.

\subsection{Numerical recipe}

\begin{enumerate}
\item Fix the leading power $p$ and the number of elements $N$;
the parent functions with indices $n=1,\dots,N+p-1$ are needed.

\smallskip

\item Compute $\beta_{nk}$ from (\ref{eq:beta}). At large $n$ it
is better not to evaluate the factorials but to run the ratio
recurrence
\begin{equation}
  \frac{\beta_{n,k+1}}{\beta_{nk}}=-\,\frac{(n-1-k)(n+2+k)}{(k+2)(k+1)},
  \qquad \beta_{n0}=\sqrt{2n+1}.
\end{equation}

\item For each $n=1,\dots,N$ solve the system (\ref{eq:sysb}) of
order $p-1$ for the weights $w^{(n)}_j$ with $w^{(n)}_0=1$. At
$p=2$ take (\ref{eq:p2}) directly instead.

\smallskip

\item Evaluate $\Psi^n(r)$ and $\sigma^n(r)$ from (\ref{eq:Psi})
with $\Phi^{n}$, $\rho^{n}$ from (\ref{eq:pair}); the Gegenbauer
polynomials are computed by the three-term recurrence \citepalias[\blue{Eq.~8.933.1}]{GR2015}
\begin{equation}
\begin{aligned}
  &(k+1)\,C^{(3/2)}_{k+1}(\xi)=(2k+3)\,\xi\,C^{(3/2)}_{k}(\xi)
  -(k+2)\,C^{(3/2)}_{k-1}(\xi),\\
  &C^{(3/2)}_0=1,\qquad C^{(3/2)}_1=3\xi .
\end{aligned}
\end{equation}

\item Take the Gram matrix from (\ref{eq:gram}); no numerical
integration is required for it. Find the expansion coefficients
from the system $\sum_n\Delta_{mn} C^n=u^m$ of Sect.~\ref{sec:p2}
(at $p=1$ simply $C^m=u^m$).

\item Control checks: (a) $r^{p}\Psi^n(r)\to c_n$ from
(\ref{eq:cn}) at large $r$; (b) $\Delta$ is banded with
half-width $p-1$, with diagonal (\ref{eq:gramdiag}), and positive
definite; (c) at $p=2$, $n=1$ the weights equal
$(1,\,-\sqrt{3/5}=-0.774597)$.
\end{enumerate}

\bsp
\label{lastpage}
\end{document}